\documentclass[fleqn,usenatbib]{mnras}

\usepackage{newtxtext,newtxmath}

\usepackage[T1]{fontenc}

\DeclareRobustCommand{\VAN}[3]{#2}
\let\VANthebibliography\thebibliography
\def\thebibliography{\DeclareRobustCommand{\VAN}[3]{##3}\VANthebibliography}

\usepackage{graphicx}	
\usepackage{amsmath}	
\usepackage{orcidlink}

\title[Neutrinos from Cyg X-3]{Microquasar Cyg X-3 -- a unique jet-wind neutrino factory?}

   \author[K.~I.~I.~Koljonen et al.]{K.~I.~I.~Koljonen$^{\orcidlink{0000-0002-9677-1533}}$,$^{1,2,3}$\thanks{E-mail: karri.koljonen@ntnu.no}
   Konstancja Satalecka,$^{2}$
   Elina J. Lindfors$^{2}$
   and Ioannis Liodakis$^{2}$
\\
   $^{1}$Institutt for Fysikk, Norwegian University of Science and Technology, H{\o}gskloreringen 5, Trondheim, 7491, Norway\\
   $^{2}$Finnish Centre for Astronomy with ESO (FINCA), University of Turku, V\"ais\"al\"antie 20, 21500 Piikki\"o, Finland \\
   $^{3}$Aalto University Mets\"ahovi Radio Observatory, PO Box 13000, FI-00076 Aalto, Finland \\
}

\date{Accepted 2023 June 16. Received 2023 May 31; in original form 2023 February 17}

\pubyear{2023}

\begin{document}
\label{firstpage}
\pagerange{\pageref{firstpage}--\pageref{lastpage}}
\maketitle

\begin{abstract}
The origin of astrophysical neutrinos is one of the most debated topics today. Perhaps the most robust evidence of neutrino counterpart comes from supermassive black holes in active galactic nuclei associated with strongly collimated outflows, or jets, that can accelerate particles to relativistic energies and produce neutrinos through hadronic interactions. Similar outflows can also be found from X-ray binaries, or `microquasars', that consist of a neutron star or a stellar-mass black hole accreting matter from a non-degenerate companion star. In some cases, these systems can accelerate particles up to GeV energies implying an efficient acceleration mechanism in their jets. Neutrino production in microquasar jets can be expected with suitable conditions and a hadronic particle population. Microquasar Cyg X-3 is a unique, short orbital period X-ray binary hosting a Wolf-Rayet companion star with a strong stellar wind. The interaction of the dense stellar wind with a relativistic jet leads to particle collisions followed by high-energy gamma-ray and potentially neutrino emission. Here, using the 10-year neutrino candidate sample of the IceCube neutrino observatory, we find that the events with the highest spatial association with Cyg X-3 occur during short-lived high-energy gamma-ray flaring periods indicating the possible astrophysical nature of these events.
\end{abstract}




\section{Introduction}

Cyg X–3 is a unique X-ray binary (XRB) \citep{giacconi67}, consisting (most likely) of a black hole and a Wolf-Rayet companion star \citep{vankerkwijk92,koljonen17} -- the only known such system in our Galaxy \citep{lommen05}. The Wolf-Rayet companion is an evolved massive star with a strong stellar wind. The short 4.8-hour orbit \citep{parsignault72} exposes the compact object to the dense wind, causing many peculiar properties \citep{koljonen18}, including high-energy (HE; $>$100 MeV) gamma-ray emission \citep{tavani09,fermi09} observed during a specific phase of an outburst event \citep{koljonen10}. 

XRBs undergo bright outburst periods that are thought to arise from a change in the rate at which matter flows from the companion star to the accretion disk. Jets in XRBs are continuously launched during the `hard' X-ray state when the X-ray emission is dominated by a hot population of electrons, typically denoted as the `corona', and are quenched in the `soft' state that is dominated by the relatively cooler emission from the accretion disk. As the system transitions from the hard to the soft state, the steady jet switches off, and a transient jet is launched, characterized by rapidly varying optically thin radio emission from blobs or shocks of plasma launched in the outflow \citep[e.g.,][]{fender99,fender04}.

The HE gamma-ray emission requires an efficient mechanism for producing relativistic particles that naturally occur in a jet. However, only a few XRBs have been seen in GeV energies. In Cyg X-3, the HE gamma-ray emission is sporadic, typically associated with the transition to/from the soft state \citep{koljonen10,koljonen18}, or occasionally during minor flaring episodes in the soft state \citep{tavani09,fermi09,corbel12,bulgarelli12}. These correspond to times when the jet is either shutting down, i.e., the system is transitioning to the soft state with quenched jet and coronal emission, or the jet is beginning to form. The HE gamma-ray emission is also orbitally modulated, placing the emitting region relatively close to the compact object \citep{fermi09}. The maximum flux occurs when the compact object is behind the companion star, where the optical depth in the line of sight through the companion stellar wind is the largest. This has provoked scenarios where the HE gamma-rays are produced by the interaction between the jet and the stellar wind either by leptonic \citep[inverse-Compton scattering of stellar photons by the relativistic electrons in the jet;][]{dubus10,zdziarski12,piano12} or hadronic processes \citep{romero03,piano12,sahakyan14}. 

In the hadronic scenario, the jet is populated with protons that collide with the protons/photons of the stellar wind. These collisions produce pions and, subsequently, gamma-rays from pion decay products that can reach very high energies (VHE; $>$100 GeV). This mechanism also produces neutrinos, which makes Cyg X-3 an attractive neutrino candidate \citep{abbasi13}. So far, Cyg X-3 has not been observed in VHE gamma-rays \citep{aleksic10,archambault13}. This could be explained by the close proximity of the emitting region to the compact object and the VHE gamma-rays having a stronger absorption probability than the lower energy ones \citep{sahakyan14}. Nevertheless, if hadronic processes produce these gamma-rays, the accompanying neutrino emission would be able to escape.

In a recent work, a neutrino signal was searched from XRBs, including Cyg X-3 \citep{abbasi22}. Using the 10-year database of neutrino events by the IceCube Observatory, they found that Cyg X-3 was the most significant source in their time-integrated search (pre- and post-trial p-values of 0.009 and 0.036, respectively). The energies of the associated events did not surpass 5 TeV, therefore, the expected neutrino event energies are low. This makes it difficult to separate the possible astrophysical neutrino candidates from background events that dominate the low energies. 

A way to suppress the background events is to use time-dependent analysis by concentrating on the times of XRB outbursts often accompanied by increased jet emission in the hard state, which would be the most likely phase of neutrino emission. \citet{abbasi22} used hard X-ray monitoring data to separate XRB outbursts from persistent/quiescent emission to increase the likelihood of the signal. However, this analysis produced a less significant signal than the time-integrated one for Cyg X-3. What sets Cyg X-3 apart from most XRBs is the unusual companion star and the short orbital period that places the jet in a dense field of stellar wind particles. \citet{koljonen18} suggested a scenario where the jet constantly evacuates a cocoon in the stellar wind during the hard state, moving the work surface further out to a less dense environment. In contrast, in the soft state, the jet switches off, allowing the wind to refill the cocoon and providing a work surface for the newly-forming jet in a much denser region when the source returns to the hard state or produces transient jet ejecta. These transitional phases are the times when we observe the HE gamma-ray emission \citep{tavani09,fermi09,corbel12,piano12,bulgarelli12}. Therefore, these times also enhance the probability of observing neutrinos with the increased rate of jet and stellar wind particle collisions. 

Considering the above, in this letter, we performed a time-dependent search during the high states of HE gamma-ray emission in Cyg X-3. We show tantalizing evidence that the HE gamma-ray emitting periods might also be periods of neutrino emission.

\section{Observations}

We utilized the IceCube Observatory \citep{aartsen17} 10-year event sample that contains track-like neutrino candidates detected between April 2008 and July 2018. The track-like signal events arise primarily from the interactions of muon neutrinos that mainly originate from cosmic rays interacting with the atmosphere but with a small fraction also from astrophysical sources. Based on the track orientation and the deposited energy, a proxy for the neutrino energy and the origin in the sky can be estimated. Details of the event selection process can be found in \citep{abbasi22}. The 10-year sample assumes a lower limit on the estimated angular uncertainty of 0.2 degrees. 

In addition, we used X-ray monitoring data from {\it Swift}/BAT\/ and obtained the daily-averaged 15--50 keV fluxes from their web interface, and weekly-averaged 0.1--100 GeV {\it Fermi}-LAT\/ detections (TS$>$4) of Cyg X-3 from the {\it Fermi}-LAT\/ light curve repository (see Data Availability). 

\section{Results}    

\begin{figure}
  \includegraphics[width=0.5\textwidth]{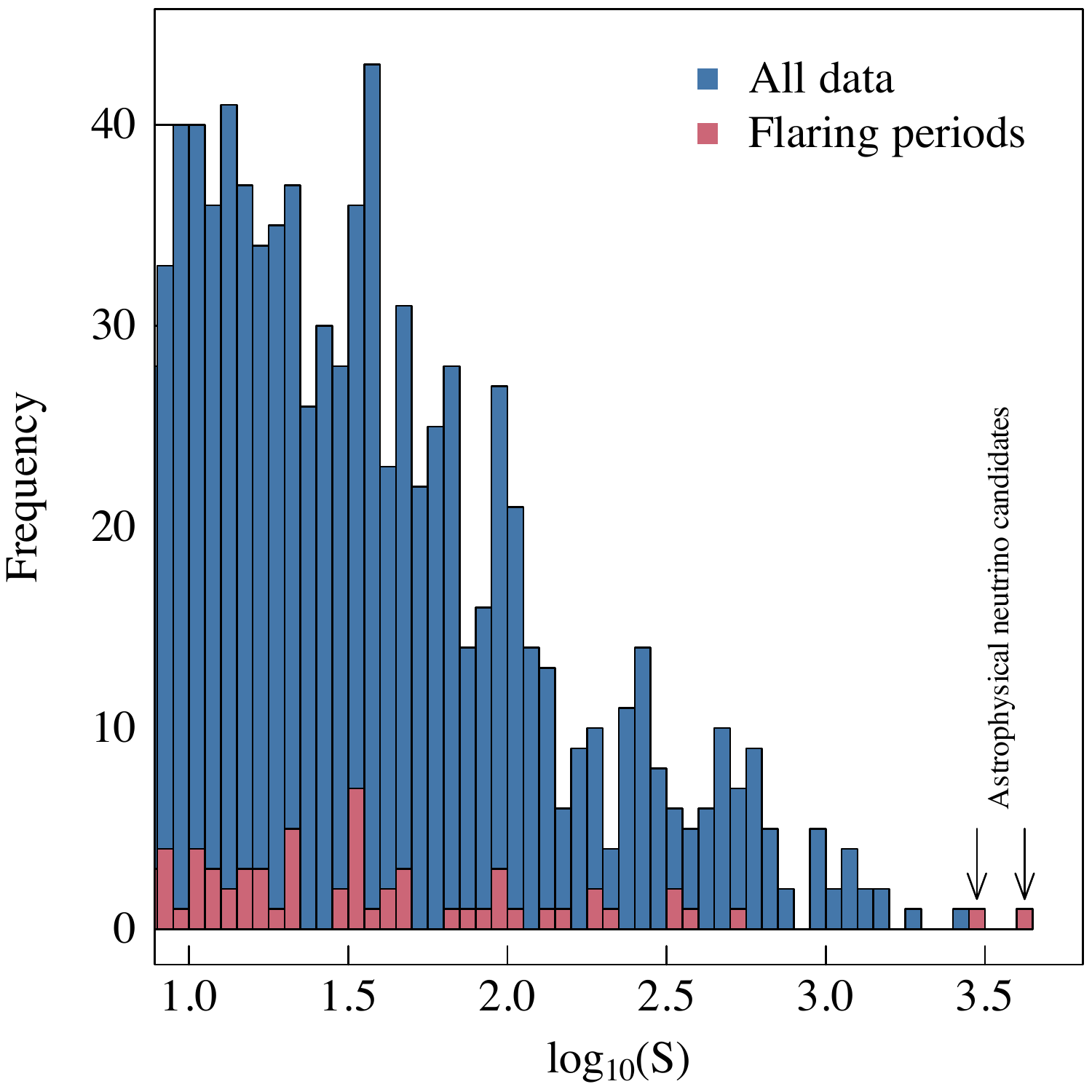}
  \caption{The spatial probability density distribution of all IceCube track-like neutrino events in the $\pm 5\deg$ band around Cyg X-3. The events occurring during the HE gamma-ray flaring periods are highlighted in red. The two astrophysical neutrino candidates are marked with arrows.}
  \label{fig:S_distribution}
\end{figure}

\begin{table}
  \scriptsize
  \centering
  \caption{The astrophysical neutrino candidate properties from Cyg X-3.}
  \label{tab:neutrino_candidates}
  \begin{tabular}{llllll}
  \\[-1.8ex]\hline
  \hline \\[-1.8ex]
  Event time & RA & Dec & Muon energy & Ang. res & Phase$^{a}$ \\ 
  (MJD) & (deg) & (deg) & (TeV) & (deg) & \\
  \hline \\[-1.8ex]
  55600.51099 & 308.127 & 40.973 & 2.09 & 0.35 & 0.8 \\
  57421.23851 & 308.478 & 41.066 & 3.31 & 0.35 & 0.4 \\
  \hline \\[-1.8ex]
  \multicolumn{6}{|p{0.85\linewidth}|}{\textit{$^{a}$} Orbital phase calculated using epheremis from \citet{antokhin19}.}
  \end{tabular}
\end{table}

\begin{figure*}
  \includegraphics[width=0.97\textwidth]{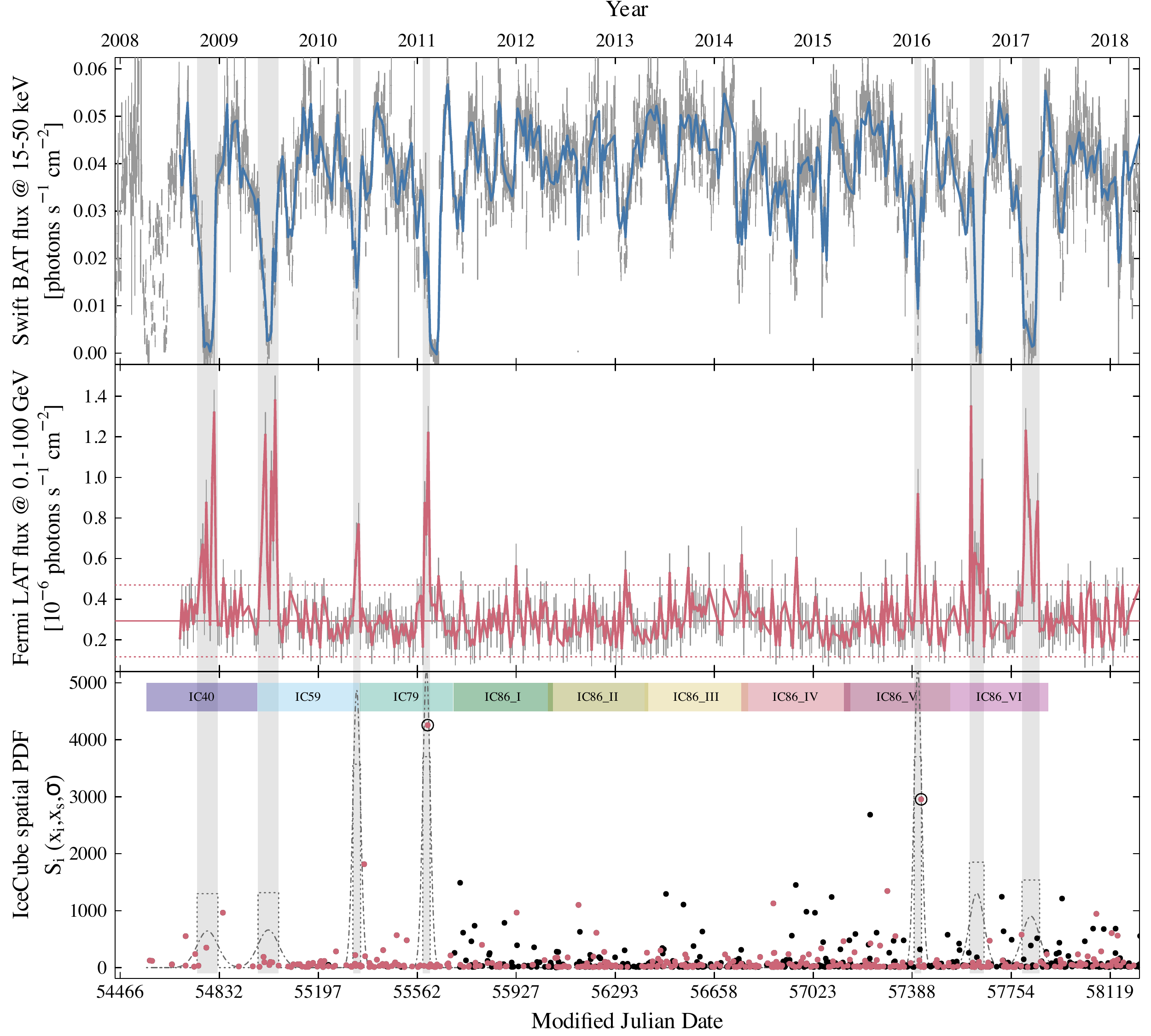}
  \caption{\textit{Top:} Daily {\it Swift}/BAT\/ 15--50 keV hard X-ray monitoring data of Cyg X-3 with a weekly average overplotted as a solid blue line. The 1$\sigma$ error bars of the daily measurement are shown in grey. \textit{Middle:} Weekly-averaged 0.1--100 GeV {\it Fermi}-LAT\/ monitoring data (TS$>$4) of Cyg X-3. HE gamma-ray flaring epochs are marked as grey areas and expanded to cover all panels. A horizontal solid line marks the average HE gamma-ray flux level outside the flaring epochs, with two standard deviations of the low-level data marked as dashed lines around the mean. The 1$\sigma$ error bars of the weekly measurements are shown in grey. \textit{Bottom:} Non-normalized spatial probability densities of IceCube events from a 10-degree band around Cyg X-3 declination. For clarity, we plot only those events with S$>$10. The events with energies higher than 1 TeV are marked in red. The astrophysical neutrino event candidates are encircled. We also plot the shapes of the top-hat (dotted lines) and gaussian (dot-dashed lines) time probability functions}.
  \label{fig:lightcurve}
\end{figure*}

To associate an IceCube track event with Cyg X-3, we make use of the spatial probability function and the expected time profile of the neutrino emission. The spatial probability function is constructed by calculating two-dimensional Gaussians in the sky with the widths based on the reconstructed angular resolution ($\sigma_{i}$) and the centroids based on the reconstructed position ($x_{i}$) of each event and recording their value at the location of the target ($x_{s}$):  

\begin{equation} \label{eq:S_values}
S_{i}(x_{i},x_{s},\sigma_{i}) = \frac{1}{2\pi\sigma_{i}^{2}} {\rm exp}\Big({-\frac{(x_{i}-x_{s})^{2}}{2\sigma_{i}^{2}}}\Big).
\end{equation}

Fig. \ref{fig:S_distribution} shows the values of the spatial probability density distributions at the location of Cyg X-3 of all IceCube track-like neutrino events and those during the HE gamma-ray flaring periods highlighted in blue and red, respectively, in the $\pm 5\deg$ band around Cyg X-3 declination. The same values are plotted in Fig. \ref{fig:lightcurve} (bottom panel) together with the hard X-ray (top panel) and the HE gamma-ray (middle panel) monitoring observations. The HE gamma-ray flaring epochs (marked as grey vertical bars in Fig. \ref{fig:lightcurve}) occur in periods of low hard X-ray emission. The two events having the strongest spatial association to Cyg X-3 (encircled points in Fig. \ref{fig:lightcurve} and arrows in Fig. \ref{fig:S_distribution}) occurred during HE gamma-ray flaring periods at the beginning of 2011 and 2016, making them intriguing candidates for astrophysical neutrinos (their properties are tabulated in Table \ref{tab:neutrino_candidates}). 

In addition, we consider two time probability functions of the expected neutrino emission. First, a top-hat profile, where the time profile is set to $T_{i}(t_{i},\sigma_{w}) = 1/\sigma_{w}$, for all events with $t_{i}$ inside the HE gamma-ray emission time interval (as shown in Fig. \ref{fig:lightcurve}, grey bands), and otherwise zero. Here, $\sigma_{w}$ is the width of a given HE gamma-ray emission time interval. Secondly, a gaussian profile, where the time profile is set to 

\begin{equation}
T_{i}(t_{i},T_{0},\sigma_{w}) = \frac{1}{\sqrt{(\pi/2)}\sigma_{w}}{\rm exp}\Big(-\frac{2(t_{i}-T_{0})^2}{\sigma_{w}^2}\Big), 
\end{equation}

\noindent where $T_{0}$ is the midpoint of a given HE gamma-ray emission time interval, and the width of the gaussian profile is considered as half of the HE gamma-ray emission time interval $\sigma_{w}$. In addition, we apply an additional weighting by the orbital phase, where the maximum weight (1) is given for orbital phases close to the superior conjuction, and the minimum weight (0.5) close to the inferior conjuction to mimic the effect of the observed high-energy gamma-ray orbital modulation \citep{fermi09,zdziarski18,prokhorov22}. The total signal probability function is then a product of the spatial and time probability functions weighted according to the orbital phase. 

To estimate the chance coincidence probability of these events being just background, we performed `pseudo experiments' using the experimental data itself. Due to the azimuthal symmetry of IceCube and a large number of background events in the data, it is possible to take a declination band around the target's declination (here, we used $\pm$5 degrees) and scramble the azimuthal components of all events in this band to give a realistic and pure background simulation. 

We simulated the background for every season separately as they have different systematic effects. The scrambling was done 10$^5$ times, and in each case, we calculated the total signal probability density of each event at the target location and recorded the event time using the two time profiles mentioned above and orbital phase weighting. These were then compared to the total signal probability densities of the actual data during the HE gamma-ray flaring windows. 

The fraction of background simulations with one event having an equal or greater total signal probability density at the location of Cyg X-3 in any HE gamma-ray flaring period as the neutrino candidate with the highest probability of association is 0.9\% (2.6$\sigma$) using the Gaussian time profile for the flares. Similarly, The fraction of background simulations with two events having an equal or greater total signal probability densities as observed in any HE gamma-ray flaring period is 0.09\%/3.3$\sigma$. Similar values were obtained with the top-hat time profile. 

In addition, we used \textsc{PSLab}\footnote{https://github.com/icecube/PSLab\_PS\_analysis}, an open source code package to perform IceCube point source analyses. Using this package, we performed time-dependent point source analysis using the top-hat time profile according to the HE gamma-ray flaring windows determined above. In this analysis, the energy of the muon event is also taken into account (assuming a fixed $\Gamma = -2.4$ spectrum). The best-fit number of signal events in this analysis is 2.6. The probability of this signal being background is 5.5\%/1.9$\sigma$. The lower probability of association as compared to the above likely arises from the soft spectrum as the bulk of the signal arises at the low, background-like energies.

\section{Discussion}

\subsection{Comparison with the expected number of neutrinos from Cyg X-3}

To estimate the expected neutrino flux emitted during the HE gamma-ray flaring periods we assume that the gamma-rays and neutrinos are produced in proton-proton collisions. This assumption is not necessarily true as a part or all the observed gamma-rays can arise from leptonic processes \citep{dubus10,zdziarski18,piano12}. We use the model of \citet{sahakyan14} that is based on the proton-proton collision process analytical estimates from \citet{kelner06} and assume a power law distributed protons with an index of $\Gamma = -2.4$ and a high energy cutoff at 100 TeV. This value is motivated by the expected cutoff in the cosmic ray energies from Galactic sources. We do not consider the proton-photon collisions here as the estimated neutrino flux at 1 TeV is two-to-three orders of magnitude smaller than in the proton-proton case \citep{baerwald13,abbasi22}. The unabsorbed HE gamma-ray fluxes range from $\sim 2 \times 10^{-12}$ erg s$^{-1}$ cm$^{-2}$ at 1 TeV down to $\sim 5 \times 10^{-13}$ erg s$^{-1}$ cm$^{-2}$ at 50 TeV. We then converted the HE gamma-ray fluxes at 1--50 TeV to neutrino fluxes using the fact that the pp-process produces an equal number of pions of all three charges and that the neutral pion decays into two gamma-ray photons for every charged pion producing a muon neutrino and antineutrino pair. For a power law spectrum with a power law index $\Gamma$, the relation of gamma-ray flux to neutrino flux is a simple factor of $2^{1-\Gamma}$ \citep[e.g.,][, their Eq. 3]{ahlers18}. We then estimated the number of neutrinos detected by IceCube with the following: 

\begin{equation}
    N_{\nu} = \int A_{\mathrm{eff}}(E)F_{\nu}(E) dE dt,  
\end{equation}

\noindent where $A_{\mathrm{eff}}(E)$ is the IceCube effective area at the declination of Cyg X-3 (41 deg) for appropriate observing seasons provided in the 10-year event sample and $F_{\nu}(E)$ is the incident neutrino flux. The integration goes through the times of the HE gamma-ray flaring periods and neutrino energies 1--50 TeV. The total accumulated gamma-ray flaring period length is 354 days during the IceCube seasons from 2008 to 2018. The estimated average number of detected $>1$ TeV neutrinos during the flaring states is 1.0. Thus, the probability of observing more than one neutrino with this average is 27\% for Poisson-distributed events.

In addition, using the upper limit on the time-integrated neutrino flux from Cyg X-3 derived in \citet{abbasi22} ($dN/dE = 8.6 (E/\mathrm{TeV})^{-2.5} \times 10^{-12} \, \mathrm{TeV}^{-1} \, \mathrm{cm}^{-2} \, \mathrm{s}^{-1}$) and assuming that all the neutrino emission takes place in the HE gamma-ray flaring periods, which then allows for a ten times higher flux, results in a maximum of 3-4 neutrinos with soft 2-3 TeV energies. Therefore, in our scenario, we are well within the upper limits derived in \citet{abbasi22}.

\subsection{The HE gamma-ray association to Cyg X-3}

The HE gamma-ray field of Cyg X-3 is complex due to nearby star formation sites, OB associations, and gamma-ray pulsars \citep{tavani09,zdziarski18}. Therefore, it is unclear whether the low-level HE gamma-ray flux can be attributed solely to Cyg X-3. In Fig. \ref{fig:lightcurve}, we delineated the phases of HE gamma-ray flares to those that rise above one standard deviation from the mean level ($>0.6 \times 10^{-6}$ photons s$^{-1}$ cm$^{-2}$ or $>3 \times 10^{-10}$ erg s$^{-1}$ cm$^{-2}$ in the 0.1--100 GeV band). These flaring events can be unequivocally attributed to Cyg X-3 due to their temporal occurrence with the phases of low flux of the hard X-ray emission (Fig. \ref{fig:lightcurve}) as well as presenting clear modulation when folded to the 4.8-hour binary orbit \citep{fermi09,zdziarski18}. This modulation has a maximum at phase 0.8, close to the superior conjunction of the binary orbit. Therefore, the probability of receiving a neutrino close to phase 0.8 is greater (as in the case of the event at MJD 55600). However, we note that the HE gamma-rays are emitted during all orbital phases \citep{prokhorov22}.  

\subsection{Contribution to the cosmic ray sea}

The origin of Galactic cosmic rays is still an open question \citep[see, e.g., a recent review by][]{gabici19}. Supernova remnants are regarded as promising sources of Galactic cosmic rays with energies up to $10^{15}$ eV -- the ''knee'' of the cosmic ray spectrum \citep[see, e.g.,][]{fermipions13,lhaaso21}. Over the last decade, space- and ground-based telescopes have revealed several new classes of Galactic gamma-ray source populations, some of which could also contribute to Galactic cosmic rays. These include for example pulsar wind nebulae \citep[e.g.,][]{hess18}, stellar clusters \citep[e.g. Cygnus cocoon;][]{fermicocoon11,hawc21}, and massive stars \citep[e.g.,][]{aharonian19}. The list also includes XRBs, which were already suggested as potential sources of cosmic rays at the beginning of the century \citep{heinz02,fender05}. While leptonic processes can partially explain the HE gamma-ray emission, neutrino detection directly implies that hadrons are accelerated to GeV-TeV energies in Cyg X-3. This implies that Cyg X-3 could contribute to its local pool of cosmic ray sea. However, due to its rare nature (we expect one black hole Wolf-Rayet binary at a given time in our galaxy; \citealt{lommen05}), the contribution at the Galaxy level is vanishingly small. 

The amount of energy contained in the protons ejected in the jet of Cyg X-3 can be estimated as $E_{p} = L_{p}t_{p} \approx 1.7 \times 10^{40}$ erg, where $L_{p} = 10^{38}$ erg/s and the proton confinement time $t_{p}$ = 0.01 $\times$ 4.8 hours \citep{sahakyan14}. As the active HE gamma-ray flaring duty cycle during the 10-year IceCube observations is approximately 10\% (354 days in 10 years), the total energy rate is $\sim 10^{39}$ erg/yr. 

\section{Conclusions}

The spatial and temporal association of neutrino emission from Cyg X-3 during HE gamma-ray flaring states has important implications, providing evidence for a new class of astrophysical neutrino emitters and showing that protons are accelerated to high energies in Cyg X-3's jet during these states. This has implications for the origin of Galactic cosmic rays and the hadron content of jets.

Microquasars and radio-loud quasars have been debated as potential sources of hadronic acceleration for decades. Given the similarity of the two systems \citep{Merloni2003,Liodakis2017}, the recent potential association of blazars with high-energy $\sim$PeV neutrinos \citep{aartsen18a} makes the scenario of lower-energy neutrinos coming from microquasars plausible. The hadron loading of the jet in microquasars can occur through the accretion of large-scale magnetic fields that mix disk matter with the jet \citep{mckinney09,cao21} or entrainment of protons from the companion star's stellar wind \citep[e.g.,][]{romero03}, both of which have a high probability of occurring in Cyg X-3 due to the short orbital separation and dense stellar wind.

The multiwavelength evolution and neutrino emission in Cyg X-3 strengthen the scenario in which the jet interacts with the dense stellar wind of the Wolf-Rayet companion. This interaction is further enhanced in the soft state when the jet is switched off and the missing ram pressure causes the wind to fill the cavity blown by the jet. As the jet begins to form again, it encounters a far denser medium than during the hard state, leading to efficient shock formation and energy dissipation through radio, gamma-ray, and neutrino emission. The recent higher duty cycle of the soft state in Cyg X-3 and corresponding times of HE gamma-ray emission may lead to more than double the neutrino detections from Cyg X-3 in upcoming IceCube data releases. 

\section*{Acknowledgements}

This project has received funding from the European Research Council (ERC) under the European Union’s Horizon 2020 research and innovation programme (grant agreement No. 101002352). K.S. and E. L. were supported by the Academy of Finland projects 317636 and 320045.

\section*{Data Availability}

The all-sky point-source muon track dataset from the IceCube Neutrino Observatory gathered between 2008 and 2018 is publicly available at http://doi.org/DOI:10.21234/sxvs-mt83. The {\it Fermi}-LAT\/ weekly light curve of Cyg X-3 can be found at https://fermi.gsfc.nasa.gov/ssc/data/access/lat/LightCurveRepository/ source.php?source\_name=4FGL\_J2032.6+4053\#. The {\it Swift}/BAT\/ daily light curve of Cyg X-3 can be found at https://swift.gsfc.nasa.gov/results/transients/CygX-3/.



\bibliographystyle{mnras}
\bibliography{bibliography} 




%
%


\bsp	
\label{lastpage}
\end{document}